\documentclass{PoS}

\title{Towards the spectrum of a GUT from gauge invariance}

\ShortTitle{Towards the spectrum of a GUT from gauge invariance}

\author{\speaker{Pascal T\"orek}\thanks{Supported by the FWF DK W1203-N16}\\
        E-mail: \email{pascal.toerek@uni-graz.at}}

\author{Axel Maas\\
        E-mail: \email{axel.maas@uni-graz.at}\\\\
         Institute of Physics, NAWI Graz, University of Graz, Universit\"atsplatz 5, 8010 Graz, Austria}

\abstract{The description of electroweak physics using perturbation theory is highly successful. Though not obvious, this is due to a subtle field-theoretical effect, the Fr\"ohlich-Morchio-Strocchi mechanism, which links the physical spectrum to that of the elementary particles. This works because of the special structure of the standard model, and it is not a priori clear whether it works for structurally different theories.

Candidates for conflicts are, e.g., grand unified theories. We study this situation in a toy model, a $SU(3)$ gauge theory with two Higgs fields and a breaking pattern $SU(3) \rightarrow SU(2) \rightarrow 1$. This mimics the weak-Higgs sector of the standard model. We determine the leading order predictions for the gauge invariant spectrum in this theory, and discuss a setup to test them using lattice gauge theory.}

\FullConference{International Symposium on Lepton Photon Interactions at High Energies\\
		17-22 August 2015\\
		University of Ljubljana, Slovenia}

\begin{document}

\section{Introduction}

One of the most important requirements in gauge theories describing particle physics is that experimentally observable states must be gauge invariant. In QCD this is realized by confinement: Only color-neutral, and thus gauge invariant, bound states of quarks and gluons are the observable objects. However, in the (electro-)weak sector of the standard model this is not obvious due to the Brout-Englert-Higgs (BEH) effect. Still, also in this case only gauge invariant objects should appear in the spectrum, which are in a non-Abelian gauge theory necessarily composite and thus describe bound states \cite{ginv,fms}. However, the experimental results are exceedingly well described utilizing the elementary, gauge-dependent states as observable particles in perturbation theory \cite{pdg}.

This apparent contradiction is resolved by the Fr\"ohlich-Morchio-Strocchi (FMS) mechanism \cite{fms}, which shows that to leading order in the Higgs fluctuations the physical states and the elementary states have the same spectrum. Given the smallness of the Higgs-fluctuations around its vev, for the parameter values leading to standard model phenomenology, this explains the success of perturbation theory. Testing the FMS mechanism theoretically requires non-perturbative methods, and has been supported using lattice simulations \cite{latwh}. It therefore appears to be the correct description of standard model physics.

However, it relies on the special structure of the standard model \cite{fms}, and it is therefore by far not clear whether it will hold beyond \cite{bsmconflict}. Understanding the situation for a model grand unified theory (GUT) is our aim here.

\section{The model and the FMS mechanism}

Our model GUT is designed to resemble at low energies only the weak-Higgs sector of the standard model. We therefore choose as a grand unified group the gauge group $SU(3)$. It will be subjected to a BEH effect twice, once at the would-be unification scale and once at the weak scale. This is implemented by using two fundamental Higgs fields, $\phi_1$ and $\phi_2$. The breaking pattern is as follows: The first Higgs field acquires a vacuum expectation value $\langle\phi_1^i\rangle = v_{1}n_{1}^i$, where $n_{1}^i=\delta^{i,3}$ is a fundamental vector. This breaks $SU(3)$ to $SU(2)$ yielding $5$ massive gauge bosons, one massive Higgs boson and $3$ massless gauge bosons. The second Higgs field breaks the $SU(2)$ symmetry to the trivial group $1$ with a vacuum expectation value $\langle\phi_{2}^i\rangle=v_{2}n_{2}^i$, with $n_{2}^i=\delta^{i,1}$. Therefore, the remaining $3$ gauge bosons also become massive and $3$ additional Higgs bosons arise.

The Lagrangian of our model is given by
\begin{equation}
\mathcal{L} = -\frac{1}{4}W_{\mu\nu}^a W^{a~\mu\nu} + \sum_{i=1}^2\left(D_\mu \phi_{i}\right)^\dagger\left(D^\mu\phi_{i}\right) - V(\phi_{1},\phi_{2}) \;,
\label{lagrange}
\end{equation}
with the field strength tensor $W_{\mu\nu}^a=\partial_{\mu} W_{\nu}^a - \partial_{\nu} W_{\mu}^a-g f^{abc}W_\mu^bW_\nu^c$ and the covariant derivative $D_\mu^{ij}=\partial_\mu\delta^{ij}-igW^{a}_{\mu}\lambda^{a~ij}/2$. The Higgs potential $V$ is chosen such that it mimics the correct phenomenology, i.e., $3$ $W$-bosons with $m_W\approx80$ GeV, one Higgs boson with $m_H\approx 126$ GeV, and $5$ gauge bosons and $3$ Higgs bosons being heavier. This can be achieved by setting
\vspace*{-0.2cm}
\begin{equation}
V(\phi_{1},\phi_{2}) = \sum_{i=1}^2 \left[-\mu_{i}^2\left(\phi_{i}^{\dagger}\phi_i\right)+\frac{\mu_{i}^2}{2v_{i}^2}\left(\phi_{i}^{\dagger}\phi_i\right)^2\right]+\alpha\left(\phi_1^\dagger\phi_2\right)\left(\phi_2^\dagger\phi_1\right)\;,
\label{potential}
\end{equation}
\vspace*{-0.5cm}
and by tuning the parameters $\mu_i$, $v_i$, $i=1,2$, and $\alpha$.

\begin{figure}[t!]
\includegraphics[width=0.49\textwidth]{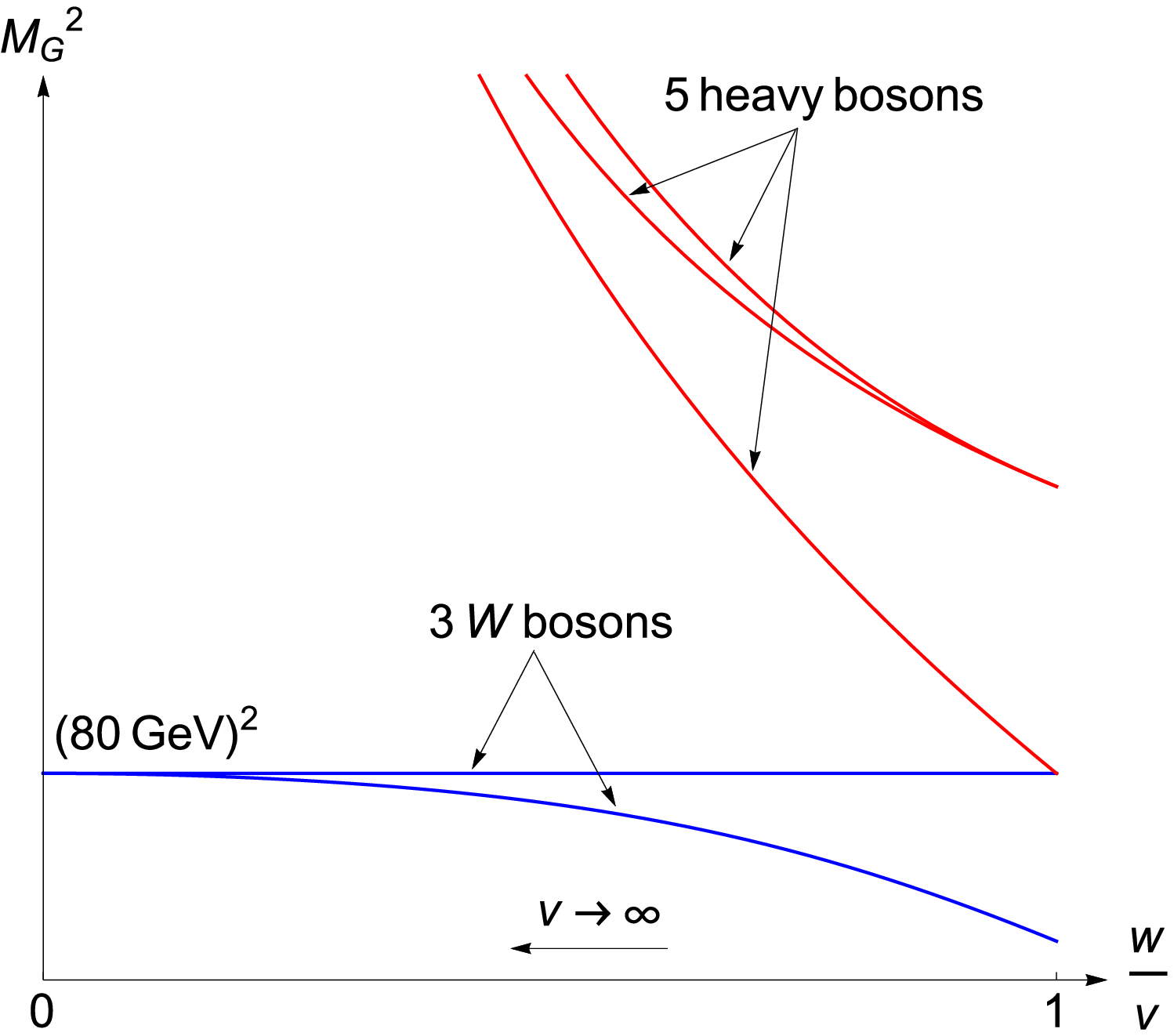}
\includegraphics[width=0.49\textwidth]{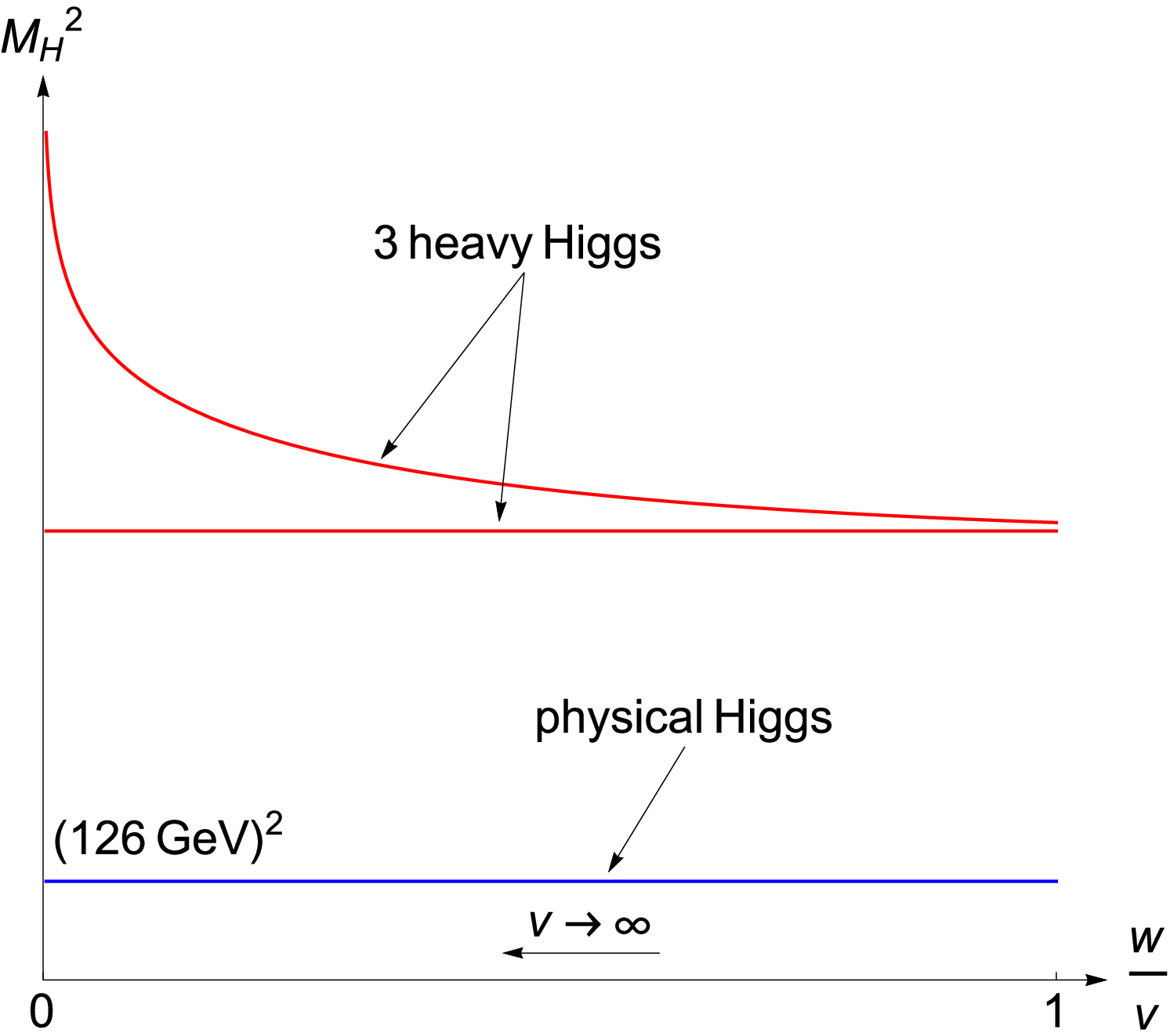}
\caption{Behavior of the gauge boson and the Higgs boson masses, $M_G^2$ and $M_H^2$, as a function of the ratio of the vacuum expectation values $v_1\equiv v$ and $v_2\equiv w$. One observes that in the limit $v\rightarrow\infty$ for a fixed value of $w$ only $3$ gauge bosons (the would-be $W$ and $Z$) and one Higgs boson (physical Higgs) have the correct phenomenological masses (blue lines). The remaining particles become very heavy (red lines).}
\label{fig1}
\end{figure}

\vspace*{-0.5cm}
Determining the mass matrices for the gauge bosons $M_G^2$ and the Higgs bosons $M_H^2$ (by expanding Eq. (\ref{lagrange}) around the vacuum expectation values $\langle\phi_i\rangle=v_in_i$, $i=1,2$) one can illustrate the behavior of those as a function of the ratio $v_2/v_1$, see Fig. \ref{fig1}. In the limit $v_1\rightarrow\infty$ (for a fixed value of $v_2$) one finds $3$ gauge bosons with masses of about $80$ GeV, i.e., the $W$ and $Z$ of the theory, and one $126$ GeV Higgs boson. The remaining $5+3$ bosons become very heavy. This is the spectrum of tree-level perturbation theory.

We now apply the FMS mechanism to the gauge invariant physical spectrum to test whether it agrees with the elementary spectrum and thus the expectations of perturbation theory. Since discrepancies are primarily expected in the vector sector \cite{bsmconflict}, we concentrate here on the $1^-$ channel. Note that there is no global symmetry in our model. Therefore, there exists only one $1^-$ channel.

A gauge invariant state with the required $J^P$ quantum numbers is created by the operator
\begin{equation}
O_\mu^{(\phi_i,\phi_j)}(x)^\dagger = \left(\phi_i^\dagger D_\mu\phi_j\right)(x)\;,\;\; i,j = 1,2\;.
\label{interpol1m}
\end{equation}
This is a Higgs-flavor tensor. The flavor symmetry is badly broken if the weak and GUT scale differ substantially.
The FMS mechanism requires to expand the bound state propagator $\langle O(x)O(y)^\dagger\rangle$ around the vacuum expectation values of the Higgs fields, i.e., $\phi_i(x) = \eta_i(x) + v_in_i$, $i=1,2$, yielding
\begin{equation}
\langle O_\mu^{(\phi_i,\phi_j)}(x)O^{\mu~(\phi_k,\phi_l)}(y)^\dagger\rangle = 
c^{ab}_{ijkl}\langle W_\mu^a(x)W^{b~\mu}(y)\rangle + \mathcal{O}(\eta W/v)\;,
\label{fms1m}
\end{equation}
where $c$ is a matrix which depends on the vacuum expectation values $\langle\phi_i\rangle$, $i=1,2$. Similar to the discussion of the FMS mechanism for the standard model \cite{fms}, we find a connection between a gauge invariant operator and the gauge variant $W$-propagator. Since the equality holds up to the order shown above, we expect the same poles on the left and right sides of (\ref{fms1m}). This implies that the masses of the states on the left-hand-side and the right-hand-side coincide.
Precisely which, and how many, states these are is determined by the matrix $c^{ab}_{ijkl}$. Its contraction with the matrix-valued $W$-propagator in (\ref{fms1m}) will create a sum over $W$-propagators in different charge directions. The appearance of any such propagator adds a mass pole which therefore is predicted to also show up in the bound state correlator on the left-hand-side. 

If there is only one Higgs field, the sum collapses to a single propagator, which belongs to the would-be $Z'$ boson in the GUT sector. Thus, in this case, the left-hand-side would have a single pole at this mass. This implies that the physical spectrum would have only a single, massive vector state. The other four states, as well as the massless states, would not appear according to this naive, i.\ e.\ without operator mixing as in \cite{fms}, application of the FMS mechanism. The difference in degeneracies is not too surprising, as there is no global symmetry, like in the standard model \cite{fms}, which could induce a physical degeneracy. The absence of the massless states is, however, unexpected.

For a theory with two Higgs fields, i.e., $i=1,2$, there are several possible correlators. Not all of them have a non-vanishing leading expansion. Those that have exhibit mass poles with masses corresponding to the would-be $Z$ as well as three of the five heavy gauge bosons, the $Z'$ and the heavier $W'^{\pm}$ doublet. Thus, once more, the naive application of the FMS mechanism yields a spectrum differing from the perturbative spectrum. 

Of course  these results hinge on the applicability of the FMS mechanism in this form. This is not only a question of structure. In the standard model case it was also found to work only in part of the parameter space \cite{latwh}. Both, the structural predictions above and the applicability of the FMS mechanism, can be tested using lattice simulation in a straight forward extension of the weak-Higgs case in the standard model \cite{latwh}. This is under way.

\section{Summary and outlook}

We have outlined possible conflicts between the gauge invariant spectrum according to the naive FMS mechanism and the perturbative, elementary spectrum for a model GUT, focusing on the vector sector. Both, in the case of a partially broken gauge group and for the fully broken case, we find possible discrepancies between the predictions from a naive application of the FMS mechanism and perturbation theory. This is quite a curious result, especially in the fully broken case. To support or refute these results requires non-perturbative calculations, which are currently conducted in form of lattice calculations. If these results should be confirmed, this could have significant implications for phenomenology using perturbation theory \cite{bsmconflict}, and therefore warrants attention.

\end{document}